\documentclass[
    aps,
    prd,
    twocolumn,
    superscriptaddress,
    floatfix
]{revtex4-1}

\usepackage[main=english]{babel}
\usepackage{amsmath}
\usepackage{amsfonts}
\usepackage{dsfont}
\usepackage{physics}
\usepackage{graphicx}

\usepackage{xcolor}
\usepackage{varwidth}
\usepackage{hyperref}
\hypersetup{linktocpage,colorlinks,citecolor={blue},pdfdisplaydoctitle=true,pdfpagemode=UseOutlines,bookmarksnumbered=true}

\newcommand*{\dgr}{\ensuremath{^{\dagger}}}

\newcommand*{\ld}{\ensuremath{l\dgr}}
\newcommand*{\rd}{\ensuremath{r\dgr}}
\newcommand*{\gauge}{\ensuremath{\mathcal{G}}}

\newcommand*{\gammain}{\ensuremath{\Gamma_{\text{in}}}}
\newcommand*{\gammainmod}{\ensuremath{\widetilde{\Gamma}_{\text{in}}}}

\newcommand*{\fid}{\ensuremath{\mathcal{A}}}
\newcommand*{\Z}[1]{\ensuremath{\mathds{Z}_{#1}}}
\newcommand*{\normsq}{\ensuremath{\abs{\Psi(\gauge)}^2}}

\newcommand*{\xarg}{\ensuremath{\vb{x}}}
\newcommand*{\gaugestate}{\ensuremath{\ket{\gauge}}}

\newcommand*{\inv}{\ensuremath{^{-1}}}
\newcommand*{\nlinks}{\ensuremath{n_{\text{links}}}}
\newcommand*{\nplaq}{\ensuremath{n_{\text{plaq}}}}

\DeclareMathOperator{\Pf}{Pf}
\DeclareMathOperator{\Adj}{Adj}

\begin{document}


\title{Variational Monte Carlo simulation with tensor networks of a pure $\Z{3}$ gauge theory in (2+1)D}


\author{Patrick Emonts}
\affiliation{Max-Planck Institute of Quantum Optics, Hans-Kopfermann-Straße 1, 85748 Garching, Germany}
\affiliation{Munich Center for Quantum Science and Technology (MCQST), Schellingstraße 4, 80799 Munich, Germany}
\author{Mari Carmen Ba\~{n}uls}
\author{Ignacio Cirac}
\affiliation{Max-Planck Institute of Quantum Optics, Hans-Kopfermann-Straße 1, 85748 Garching, Germany}
\affiliation{Munich Center for Quantum Science and Technology (MCQST), Schellingstraße 4, 80799 Munich, Germany}
\author{Erez Zohar}
\affiliation{Racah Institute of Physics, The Hebrew University of Jerusalem, Givat Ram, Jerusalem 91904, Israel}


\date{\today}

\begin{abstract}
Variational minimization of tensor network states enables the exploration of low energy states of lattice gauge theories.
However, the exact numerical evaluation of high-dimensional tensor network states remains challenging in general.
In~[E. Zohar, J. I. Cirac, Phys. Rev. D \textbf{97}, 034510 (2018)] it was shown how, by combining gauged Gaussian projected entangled pair states with a variational Monte Carlo procedure, it is possible to efficiently compute physical observables.
In this paper we demonstrate how this approach can be used to investigate numerically the ground state of a lattice gauge theory.  
More concretely, we explicitly carry out the variational Monte Carlo procedure based on such contraction methods for a pure gauge Kogut-Susskind Hamiltonian with a $\Z{3}$ gauge field in two spatial dimensions.
This is a first proof of principle to the method, which provides an inherent way to increase the number of variational parameters and can be readily extended to systems with physical fermions.
\end{abstract}

\maketitle

\section{Introduction}
Tensor network states, especially matrix product states (MPSs), have changed our understanding of solid state systems dramatically.
Describing states with an area-law entanglement, i.e. ground states of local, gapped Hamiltonians~\cite{orus_practical_2014,cirac_renormalization_2009}, MPSs provide an ansatz class for a wide range of problems due to their favorable numerical scaling.
Instead of an exponential scaling, MPS algorithms scale polynomially with the system size.
The computational power in combination with a solid analytical understanding allowed a variety of applications, including ground state searches~\cite{white_density_1992,schollwock_density-matrix_2011} and the description of dynamics of many-body systems.
Similar studies have been performed with tensor networks in two spatial dimensions, projected entangled pair states (PEPSs)~\cite{corboz_simulation_2010}.

Motivated by the success of tensor networks in condensed matter physics, such methods have been generalized and applied to particle physics problems too, in particular to lattice gauge theories (LGTs)~\cite{banuls_review_2020}.
Gauge theories appear in many fundamental physical contexts, e.g. the standard model of particle physics, where gauge fields act as force carriers.
In particular, it includes quantum chromodynamics (QCD), the theory of the strong nuclear force, which, as a non-Abelian gauge theory \cite{peskin_introduction_1995} has a running coupling.
In QCD, asymptotic freedom \cite{gross_asymptotically_1973} gives rise to asymptotically weak couplings for high energy scales (e.g. collider experiments), and therefore perturbation theory could be used in these physical regimes. 
On the other hand, low energy QCD is a strongly coupled model, requiring nonperturbative treatment.

One approach to regimes where nonperturbative methods break down is lattice gauge theories.
They provide a gauge invariant regularization of gauge theories, discretizing either spacetime~\cite{wilson_confinement_1974} or only space (leaving time continuous)~\cite{kogut_hamiltonian_1975}.
Simulations based on hybrid Monte Carlo~\cite{duane_hybrid_1987,flag_working_group_review_2014} have given many interesting insights into the physics in the nonperturbative regime.
While having been extremely successful and fruitful for static studies (such as studies of the hadronic spectrum), this method faces two major difficulties. 
First, the inability to directly observe time dependent phenomena in Wick-rotated, Euclidean spacetimes, as done in this context; the second is the well-known sign problem \cite{troyer_computational_2005} which appears in scenarios with finite fermionic chemical potential, where the statistical interpretation allowing one to perform Monte Carlo sampling breaks down, blocking the way to important phases of the QCD phase diagram \cite{fukushima_phase_2011}.

In (1+1)D, MPSs have been very successful describing LGTs (see Ref.~\cite{banuls_review_2020} and references therein).
In higher dimensions, MPSs are generalized to PEPSs, whose contraction is in general very costly. 
This hinders the application of variational PEPS algorithms in higher dimensions, although state of the art algorithms can handle all the terms in a gauge theory~\cite{schulz_breakdown_2012} and a first numerical study for a pure gauge theory has been recently presented in~\cite{robaina_simulating_2020}.
Earlier numerical studies used less general tensor networks for two-dimensional lattice gauge theories, either purely gauge~\cite{tagliacozzo_entanglement_2011} or including fermions~\cite{felser_two-dimensional_2019}.
In contrast, analytical approaches have developed faster, with the formulation of gauge invariant pure gauge PEPSs~\cite{tagliacozzo_tensor_2014}, and more general gauging mechanisms including matter for arbitrarily dimensional PEPSs~\cite{haegeman_gauging_2015,zohar_building_2016}.

In these works, the global symmetry of a matter-only PEPS is lifted to a local one by introducing a gauge field, in a way analogous to minimal coupling.
The latter gauging method has been used for the construction of gauged Gaussian fermionic PEPSs~\cite{zohar_fermionic_2015,zohar_projected_2016}, where the matter state to be gauged is a free (Gaussian) fermionic state, in a manner analogous to minimal coupling of a Hamiltonian~\cite{emonts_gauss_2020}. 
The restriction to this subclass of PEPSs enables the efficient contraction of the states with Monte Carlo techniques~\cite{zohar_combining_2018}.
Since the sampling probability of the algorithm depends only on the norm of the state, the Monte Carlo algorithm cannot suffer from the sign problem.
Furthermore, the construction allows for a natural and efficient extension to higher bond dimensions which is numerically very expensive in general PEPS calculations.
However, until now, these states have only been used to compute observables of toy models -- either exact contractions, showing relevant physical behavior~\cite{zohar_fermionic_2015,zohar_projected_2016} or a demonstration of the feasibility of the Monte Carlo contraction of the PEPS, but for given states, without variational techniques~\cite{zohar_combining_2018}. 

The next step, required for demonstrating the credibility and feasibility of the method, is the actual variation (energy minimization) procedure of a real lattice gauge theory Hamiltonian: a numerical verification that the such ansatz states can converge to true ground states.  
In this paper, we present the application of fermionic gauged Gaussian PEPSs~\cite{zohar_fermionic_2015,zohar_building_2016,zohar_projected_2016,zohar_combining_2018} as ansatz states in a variational Monte Carlo (VMC) procedure~\cite{sorella_generalized_2001,sorella_wave_2005,sandvik_variational_2007}.
We apply the algorithm to a Hamiltonian pure $\Z{3}$ gauge theory~\cite{horn_hamiltonian_1979} and make explicit use of the possibility to extend the ansatz efficiently by adding more layers of virtual parameters.

The $\Z{3}$ theory is a relatively simple (2+1)D theory, but it is known to exhibit a (first-order) phase transition between a confining and nonconfining phase, and thus constitutes a nontrivial testbench for the ansatz~\cite{bhanot_phase_1980}.
Furthermore, extensive Monte Carlo studies have been performed on $\Z{N}$ theories, which allow us to benchmark our results against known results~\cite{blote_first-order_1979}.
Our goal is to demonstrate the expressibility of the ansatz presented in Ref.~\cite{zohar_combining_2018} and how it can be applied to study gauge theories.  
Adding more layers to the construction is essential to improve convergence, especially in the low coupling regime of the theory.
However, precisely locating the phase transition remains challenging, even with an increased number of layers.
The main obstacle is the expensive evaluation of a Pfaffian that appears in the calculation of the electric energy.
Thus, it has to be calculated in every Monte Carlo step during the energy minimization.

The rest of the manuscript is structured as follows: 
In Secs.~\ref{sec:hilbert_space} and~\ref{sec:peps_construction}, we introduce $\Z{N}$ gauge theories and construct our ansatz states.
These states are minimized with the numerical methods described in Sec.~\ref{sec:comp_evaluation}.
The numerical results are presented in Sec.~\ref{sec:results}.
Finally, we conclude in Sec.~\ref{sec:conclusion}.

\section{Hilbert space of Abelian lattice gauge theories\label{sec:hilbert_space}}
\begin{figure}
  \centering
  \includegraphics[]{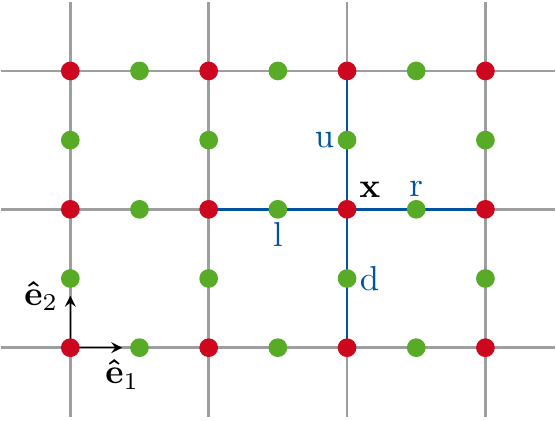}
  \caption{
    Arrangement of fermions and gauge fields in a lattice gauge theory. 
    Fermions are indicated in red, gauge fields are shown in green.
    The convention for labeling links around a vertex $\xarg$ is indicated in blue.
  }
  \label{fig:lattice}
\end{figure}

In a Hamiltonian lattice gauge theory, space is discretized and represented on a lattice while time remains continuous~\cite{kogut_hamiltonian_1975}.
This is in contrast to the action formulation, where both space and time are discretized~\cite{wilson_confinement_1974}.
The (fermionic) matter of the theory resides on the vertices $\xarg$ of a lattice, and the interactions are mediated by  gauge fields, whose quantum Hilbert spaces reside on the links (compare Fig.~\ref{fig:lattice}).
In the following, we will focus on Abelian lattice gauge theories with finite gauge groups ($\Z{N}$), without dynamical matter, i.e. pure gauge theories.
We will consider a two-dimensional $L\times L$  lattice with periodic boundary conditions.
Thus, the only degrees of freedom of the theory reside on the links.

One problem of numerically simulating a lattice gauge theories with compact Lie groups [even the Abelian $U(1)$] is the infinite dimension of the Hilbert spaces on the links.
This can be approached by truncating the local Hilbert spaces, either by introducing a cutoff to the electric field, allowing one to restore the full theory by extending the cutoff~\cite{zohar_fermionic_2015} or integrating over an extra dimension~\cite{horn_finite_1981,orland_lattice_1990,chandrasekharan_quantum_1997}, or by sampling group elements~\cite{horn_hamiltonian_1979} from the gauge group, which form a subgroup.
Due to the construction of our states (see Sec.~\ref{sec:peps_construction}), we chose the second approach, i.e. instead of simulating the full $U(1)$ theory, we consider a $\Z{N}$ subgroup that serves as an approximation for $U(1)$.
As described in Ref.~\cite{horn_hamiltonian_1979}, the $N\to\infty$ limit of $\Z{N}$   reproduces  $U(1)$, and hence $\Z{N}$ lattice gauge theories flow, in the large $N$ limit, to compact QED \cite{kogut_introduction_1979}, a lattice gauge theory with $U(1)$ symmetry.

\begin{figure}
  \centering
  \includegraphics[]{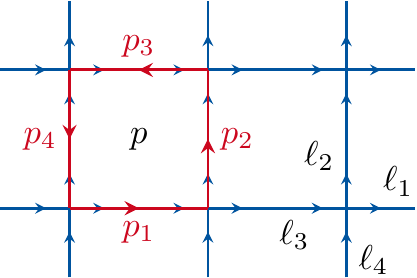}
  \caption{
    Convention for labeling the links of a plaquette. 
    The red arrows indicate the orientation of the plaquette.
    The blue arrows show the convention for the calculation of a divergence on the lattice.
  }
  \label{fig:plaquette}
\end{figure}

We write the Hamiltonian of a pure $\Z{N}$ gauge theory as 
\begin{align}
  H&=H_E+H_B\nonumber\\
  &=\frac{g^2}{2}\sum_\ell\left[2-(P_\ell+P_\ell\dgr)\right]\nonumber\\
  &\phantom{=}+\frac{1}{2g^2}\sum_p\left[2-(Q_{p_1}\dgr Q_{p_2}\dgr Q_{p_3}Q_{p_4}+\text{H.c.})\right],
  \label{eq:hamilton_ZN}
\end{align}
where $\ell=(\xarg,i)$ is a link on the lattice emanating from vertex $\xarg$ horizontally($i=\vu{e}_1$) or vertically ($i=\vu{e}_2$) and $p$ is a plaquette~\cite{horn_hamiltonian_1979}.
The indices $p_j$ refer to one of the four links of one plaquette as indicated in Fig.~\ref{fig:plaquette}.
The terms $H_E$ and $H_B$ are referred to as electric and magnetic part of the Hamiltonian, respectively~\cite{kogut_hamiltonian_1975}.

The operators in~\eqref{eq:hamilton_ZN} obey the $\Z{N}$ algebra given by
\begin{equation}
  \label{eq:ZN_algebra}
  \begin{aligned}
    &P_\ell^N=Q_\ell^N=1&& P_\ell\dgr P_\ell=Q_\ell\dgr Q_\ell=1\\
    &P_\ell\dgr Q_\ell P_\ell  =e^{i\delta Q_\ell} && \delta=\frac{2\pi}{N}.
  \end{aligned}
\end{equation}
Operators that act on different links commute with each other.

The Hamiltonian~\eqref{eq:hamilton_ZN} is invariant under the action of the local unitary operators
\begin{align}
  \Theta(\xarg)=P_{\xarg,r}P_{\xarg,u}P\dgr_{\xarg-\vu{e}_1,r}P\dgr_{\xarg-\vu{e}_2,u}.
  \label{eq:gauge_inv_p}
\end{align}
The links are addressed according to their vertex $\xarg$ and their direction right ($r$) or up ($u$).
This local gauge invariance implies that $\Theta(\xarg)$ commutes with the Hamiltonian on each site 
\begin{align}
  \comm{\Theta(\xarg)}{H}=0\quad\forall\,\xarg.
\end{align}
Due to the generators of local symmetry (given in~\eqref{eq:gauge_inv_p}), we know that the physical states of the system obey the symmetry
\begin{align}
  \Theta(\xarg)\ket{\Psi}=\ket{\Psi}\quad\forall\,\xarg.
  \label{eq:application_gauss_law}
\end{align}
Equation~\eqref{eq:application_gauss_law} holds since we do not consider static charges in this work.

Given the $\Z{N}$ group, we define a set of group element states $\ket{q(\ell)}$ labeled by integers $q=0,...,N-1$,  which span the local gauge field Hilbert space on link $\ell$.
They correspond to group elements with the discrete angles $\phi(l)=q\delta$ [$\delta$ is defined in \eqref{eq:ZN_algebra}].
The group element states form an orthonormal basis for the local Hilbert space $\bra{q}\ket{q'}=\delta_{q,q'}$.

These states are eigenstates of the $Q$ operators, with
\begin{equation}
    Q\ket{q}=e^{i\delta q}\ket{q}.
\end{equation}
They are lowered by the $P$ operators, periodically:
\begin{equation}
    P\ket{q}=\ket{q-1}.
\end{equation}

\section{PEPS construction with Abelian symmetry\label{sec:peps_construction}}

Products of local group element states define the configuration of gauge fields on the lattice. 
Such product states, $\gaugestate=\otimes_{\ell}\ket{q(\ell)}$ form an orthonormal basis, using which we can expand every state in the gauge field Hilbert space:
\begin{align}
  \ket{\Psi}=\sum_{\mathcal{G}} \Psi(\gauge)\gaugestate,
  \label{eq:ansatz_state}
\end{align}
where the sum runs over all possible gauge field configurations on the links and $\Psi(\gauge)$ is a gauge field dependent wave function of the configuration $\gauge$. 
This expression is a special case of the more general formulation presented in~\cite{zohar_combining_2018}, where $\Psi(\gauge)$ can be a quantum state of the dynamical (fermionic) matter, $\ket{\Psi(\gauge)}$, instead of the wave function we have in our current pure gauge case.

Not every state that can be expressed with~\eqref{eq:ansatz_state} is physically relevant, i.e. fulfills the local symmetry~\eqref{eq:gauge_inv_p}.
Thus, the wave function $\Psi(\gauge)$ has to be chosen such that the full state $\ket{\Psi}$ obeys the correct symmetries.
Additionally, the state that we pick should allow for efficient numerical calculations of observables and gradients. 
Following the general construction in~\cite{zohar_combining_2018}, we choose a gauged Gaussian projected entangled pair state (GGPEPS) as an ansatz.
For details and further motivation, we refer to Refs.~\cite{emonts_gauss_2020,zohar_fermionic_2015}.

\subsection{Construction with a single layer}
Following the idea of a PEPS construction, we build the GGPEPS out of local constituents which help us to impose the symmetry. 
The local parts are entangled during the construction to form the final wave function.

The elementary building blocks for the wave function are auxiliary (or virtual) fermionic modes that are attached to each outgoing and ingoing leg of each  vertex of the lattice.
They are chosen to be fermionic to enable a consistent coupling to fermionic matter which obeys the correct statistics~\cite{zohar_combining_2018}.
Although, for the description of a pure gauge theory, the coupling to matter is not necessary.

\begin{figure}[h]
  \centering
  \includegraphics{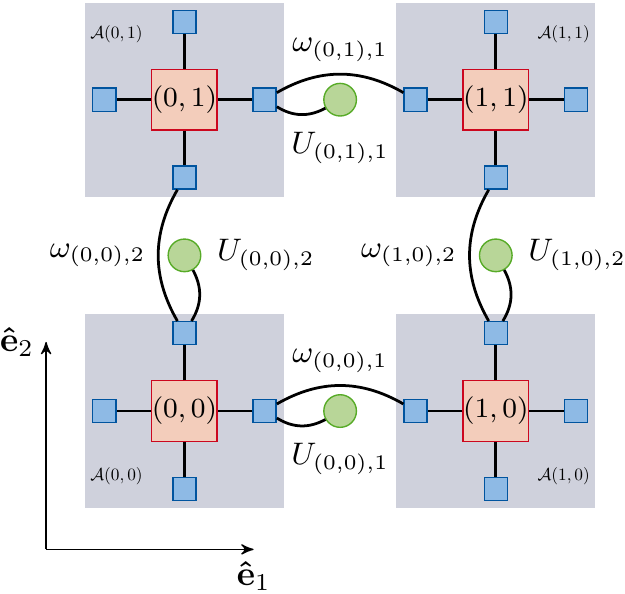}
  \caption{
    Illustration of the state's construction. 
    The interior of the grey squares is created by the fiducial operator $\fid$.
    Blue squares mark virtual modes in the different directions.
    The bent lines between the virtual modes illustrate the unnormalized projectors $\omega$.
    The gauge fields on the links between the sites are depicted as green circles.
    Their coupling to the virtual respective modes is shown as bent lines as well.
  }
  \label{fig:state_local}
\end{figure}

The construction of a GGPEPS consists of three essential parts (cmp. Fig.~\ref{fig:state_local}).
First, the \emph{fiducial operators} $\fid(\xarg)$ create virtual fermionic states out of the modes associated with each site. 
They are constructed in a way that guarantees virtual gauge invariance (used in general PEPS constructions for imposing global symmetries).
This step of the construction can be readily extended to include more virtual fermions, in a similar spirit that the bond dimension of a PEPS can be increased.
The details of the construction with multiple layers are given below.
Then, some of the virtual modes on each site are rotated with respect to the physical gauge fields of the theory, in a particular way that lifts the virtual symmetries to physical ones \cite{zohar_combining_2018}.
This is done by gauging operators $\mathcal{U}_{\gauge}$ acting  on the virtual fermions and controlled by the gauge field configuration. 
Finally, the pairs of virtual fermionic modes on the two sides of each link are projected onto maximally entangled states by projection operators $\omega_\ell$. That contracts the state from its local constituents and introduces correlations to the state. 

The wave function can thus be written as
\begin{align}
  \Psi(\mathcal{G})=\bra{\Omega_v}\prod_{\ell}\omega_\ell\prod_{\ell}\mathcal{U}_{\mathcal{G}}(\ell)\prod_{\xarg}\fid(\xarg)\ket{\Omega_v},
  \label{eq:fermionic_state_gauged}
\end{align}
where the products are over all links $\ell$ of the lattice and $\ket{\Omega_v}$ is the fermionic Fock vacuum.
In the following, we will treat the three main components of the construction $\fid$, $\mathcal{U}_{\gauge}$, and $\omega$ in more detail, and see how to make sure that $\Psi(\mathcal{G})$ obeys the right symmetry properties.
Furthermore, aiming at an efficient computation of the wave function, we would like it to be Gaussian, and thus all its constituents will be Gaussian too.

On each vertex~$\xarg$ of the two-dimensional lattice, we define eight virtual fermionic modes, two associated to each leg - left, right, up and down.
On each leg we label the two modes by $\pm$, and sort them into two groups: $a_i=\left\{l_{+},r_{-},u_{-},d_{+}\right\}$ (which we call the negative modes) and $b_i=\left\{ l_{-},r_{+},u_{+},d_{-} \right\}$ (positive modes).
The modes obey the Dirac anticommutation relation $\acomm{c(\xarg)}{c\dgr(\vb{y})}=\delta_{\xarg,\vb{y}}$ and $\acomm{c(\xarg)}{c(\vb{y})}=\acomm{c\dgr(\xarg)}{c\dgr(\vb{y})}=0$, where $\xarg,\vb{y}$ are vertices on the lattice and $c$ is a fermionic mode.

We define the virtual electric fields
\begin{align}
  E_0(\xarg,k)=(-1)^{\xarg}(k_+\dgr(\xarg)k_+(\xarg)+k_-\dgr(\xarg)k_-(\xarg))
  \label{eq:staggered_el_field}
\end{align}
with $k\in\{r,l,u,d\}$ as well as the generator of the gauge transformation on the virtual degrees of freedom,
\begin{align}
  G_0(\xarg)=E_0(\xarg,r)+E_0(\xarg,u)-E_0(\xarg,l)-E_0(\xarg,d).
  \label{eq:gauss_law_peps}
\end{align}
This can be seen as a version of a Gauss law operator: the divergence of the virtual electric fields at the vertex.
The staggering  is introduced to accommodate the general case with physical fermions \cite{zohar_combining_2018} (aiming at the problem of physical fermion doubling \cite{susskind_lattice_1977} which we do not encounter in the pure gauge case). 
It is taken care of already on the level of electric fields [cmp.~\eqref{eq:staggered_el_field}] and thus the rest of the equations can be stated without explicit reference to staggering.

The fiducial operator $\fid(\xarg)$ which creates the modes out of the vacuum has to be Gaussian, and be invariant under transformation generated by $G_0(\xarg)$. 
Hence, it is given by \cite{zohar_fermionic_2015,emonts_gauss_2020}
\begin{align}
  \fid(\xarg)= \exp\left(\sum_{ij}T_{ij}a^{\dagger}_i(\xarg)b^{\dagger}_j(\xarg)\right),
  \label{eq:def_fid_op}
\end{align}
where $T_{ij}$ is a $4\times 4$ matrix containing all parameters of the ansatz. 
$\fid$ is a Gaussian operator by construction, and one can easily inspect that since positive modes are only coupled to negative ones, the symmetry property 
\begin{align}
  \exp(i\alpha G_0(\xarg))\fid(\xarg)\exp(-i\alpha G_0(\xarg))=\fid(\xarg).
  \label{eq:invariance_fid_operator}
\end{align}
is satisfied for every angle $\alpha$, hence forming a $U(1)$ parameterization. 
As such, it holds also for the $\Z{N}$ cases, with a discrete choice of angles.
Due to other symmetry considerations (e.g. lattice rotation invariance), only two independent parameters in $T_{ij}$ of initially sixteen remain, $y$ and $z$. 
They couple different modes in a given vertex:
$y$ couples right(up) and left(down) modes in a vertex, $z$ couples modes that are building corners, e.g. right and up modes.
The exact form of $T$ and a motivation of the symmetries can be found in Appendix~\ref{sec:app_tmat}.

For now, we will formulate the ansatz with eight virtual fermions per vertex.
One set of eight virtual fermions is referred to as one \emph{layer}.
In a second step, we will enlarge the number of variational parameters by adding more layers, i.e more virtual fermions to the links.
Each layer gets an independent set of parameters $y$ and $z$.
Increasing the number of layers is the analogue to increasing the virtual bond dimension in a non-fermionic PEPS.

In a second step, we entangle the virtual fermions on the links with physical gauge fields on the links.
The gauging operator for a given gauge field configuration $\mathcal{G}$ takes the form
\begin{align}
  \mathcal{U}_{\mathcal{G}}\left(\ell\right)=
  \begin{cases}
    e^{i\left(-1\right)^{\xarg}q\left(\ell\right)\delta E_{0}(\vb{x},r)} & \ell\:\mathrm{horizontal}\\
    e^{i\left(-1\right)^{\xarg}q\left(\ell\right)\delta E_{0}(\vb{x},u)} & \ell\:\mathrm{vertical}.
  \end{cases}
\end{align}
where $q(\ell)$ parameterizes the group element on the link $\ell$ in the configuration $\mathcal{G}$.
The local gauge transformation changes only the modes pointing up and right.
Modifying the left and bottom modes as well would undo the gauge transformation due to the staggering.
For a detailed overview of the gauging procedure in terms of PEPS operators, i.e. in graphical notation, we refer to Refs.~\cite{zohar_building_2016,zohar_fermionic_2015,emonts_gauss_2020}.

In order to create more than a product state, we project the virtual, fermionic modes adjacent to each link onto maximally entangled states.
The unnormalized projectors 
\begin{align}
  \omega_{\xarg,1}&=\nonumber\\
  &\exp\left(l_{+}^{\dagger}\left(\xarg+\vu{e}_{1}\right)r_{-}^{\dagger}\left(\xarg\right)+l_{-}^{\dagger}\left(\xarg+\vu{e}_{1}\right)r_{+}^{\dagger}\left(\xarg\right)\right)\Omega_{\ell}\times\nonumber\\
  &\times\exp\left(r_{-}\left(\xarg\right)l_{+}\left(\xarg+\vu{e}_{1}\right)+r_{+}\left(\xarg\right)l_{-}\left(\xarg+\vu{e}_{1}\right)\right)\\
  \omega_{\xarg,2}&=\nonumber\\
  &\exp\left(u_{+}^{\dagger}\left(\xarg\right)d_{-}^{\dagger}\left(\xarg+\vu{e}_{2}\right)+u_{-}^{\dagger}\left(\xarg\right)d_{+}^{\dagger}\left(\xarg+\vu{e}_{2}\right)\right)\Omega_{\ell}\times\nonumber\\
  &\times\exp\left(d_{-}\left(\xarg+\vu{e}_{2}\right)u_{+}\left(\xarg\right)+d_{+}\left(\xarg+\vu{e}_{2}\right)u_{-}\left(\xarg\right)\right),
\end{align} 
connect the left(upper) and right(lower) modes of neighboring sites.
Here, $\Omega_{\ell}$ is the projector to the virtual vacuum on link $\ell$ and  $\vu{e}_i$ is the unit vector in direction $i$.
Similar to the fiducial operators $\fid$, the projectors $\omega$ are Gaussian and commute among each other since they are products of fermionic modes on different links.
The projectors link the virtual modes of one site with the virtual modes of the next site in the horizontal and the vertical direction, respectively.
It is essential that the projectors are unnormalized since the norm of a state will serve as a transition probability between different gauge field configurations later.

Combining $\fid$, $\omega$, and $\mathcal{U}_{\mathcal{G}}$, we get the wave function in Eq.~\eqref{eq:fermionic_state_gauged}.
Now, we can show that the construction is indeed gauge invariant and fulfills~\eqref{eq:application_gauss_law}.
We act with $\Theta(\xarg)$ on $\ket{\Psi}$ explicitly, on some given vertex $\mathbf{x}$:
\begin{widetext}
  \begin{align}
    \begin{aligned}
      \Theta(\xarg)\ket{\Psi}&=\sum_{\gauge}\Psi(\gauge)P_{\xarg,u} P_{\xarg,r} P\dgr_{\xarg-\vu{e}_1,r}P\dgr_{\xarg-\vu{e}_2,u}\ket{\gauge}\\
      &=\sum_{\gauge}\Psi(\gauge) \ket{q(\ell_1)-1,q(\ell_2)-1,q(\ell_3)+1,q(\ell_4)+1}\otimes\ket{\tilde{q}}\\
      &=\sum_{\gauge}\Psi(\underbrace{q(\ell_1)+1,q(\ell_2)+1,q(\ell_3)-1,q(\ell_4)-1,\tilde{q}}_{\equiv\gauge'})\ket{\gauge},
      \end{aligned}
    \label{eq:proof_invariance}
  \end{align}
\end{widetext}
where $\tilde{q}$ are all gauge fields that are not affected by the gauge transformation, i.e. that are not adjacent to $\xarg$.
To shorten notation, we named the different links according to the labels defined in Fig.~\ref{fig:plaquette}.
The third line is linked to the second one by a change of variables in $q$.
The gauge invariance holds if $\Psi(\gauge)=\Psi(\gauge')$.
We can write the wave function $\Psi(\gauge')$ as
\begin{widetext}
  \begin{align}
    \Psi(\gauge')=&\bra{\Omega_v}\prod_\ell \omega_{\ell} \prod_{\tilde{\ell}}\mathcal{U}_{\gauge}(\tilde{\ell})e^{\pm i\delta(q_1+1)E_0(\xarg,r)}e^{\pm i\delta(q_2+1)E_0(\xarg,u)}e^{\mp i\delta(q_3-1)E_0(\xarg-\vu{e}_1,r)}e^{\mp i\delta(q_4-1)E_0(\xarg-\vu{e}_2,u)}\prod_{\xarg} \fid(\xarg)\ket{\Omega_v}\nonumber\\
    =&\bra{\Omega_v}\prod_\ell \omega_{\ell} \prod_{\tilde{\ell}}\mathcal{U}_{\gauge}(\tilde{\ell})e^{\pm i\delta (E_0(\xarg,r)+E_0(\xarg,u)-E_0(\xarg,l)-E_0(\xarg,d))}\prod_{\xarg}\fid(\xarg)\ket{\Omega_v}\nonumber\\
    =&\Psi(\gauge),
  \end{align}
\end{widetext}
where $\tilde{\ell}$ are all links that are unaffected by the gauge transformation and $\Omega_v$ is the vacuum of all virtual modes.
The notation of multiple signs shows the transformation for an even (top sign) and an odd (bottom sign) vertex at the same time.
We used the invariance of the fiducial operator~\eqref{eq:invariance_fid_operator} at the last line.
In order to transform the virtual electric field from the adjacent vertices $\xarg-\vu{e}_1$ and $\xarg-\vu{e}_2$ to vertex $\xarg$, we use the invariance of the projectors $\omega$:
\begin{align}
  \begin{aligned}
  \omega_{\xarg-\vu{e}_1,1} e^{i\delta E_0(\xarg-\vu{e}_1,r)}=\omega_{\xarg-\vu{e}_1,1} e^{-i\delta E_0(\xarg,l)}\\
  \omega_{\xarg-\vu{e}_2,2} e^{i\delta E_0(\xarg-\vu{e}_2,u)}=\omega_{\xarg-\vu{e}_2,2} e^{-i\delta E_0(\xarg,d)}
  \end{aligned}
  \label{eq:invariance_omega}
\end{align}

All operators employed in the construction ($\fid$, $\omega$, and $\mathcal{U}_{\mathcal{G}}$) are Gaussian operators.
Since products of Gaussian operators are still Gaussian~\cite{bravyi_lagrangian_2005}, the wave function $\Psi(\gauge)$ can be efficiently described with covariance matrices.
As detailed in \cite{zohar_combining_2018}, there are multiple ways of combining the operators to covariance matrices.
We choose to group the gauging operators and the projectors together into $\gammain(\gauge)$, a covariance matrix that depends on the gauge.
The fiducial operators are summarized in a second covariance matrix $D$.
The relation between the covariance matrices and the gauged ansatz state can be summarized as
\begin{align}
  \Psi(\mathcal{G})=\underbrace{\bra{\Omega_v}\prod_{\xarg}\omega(\xarg)\prod_{\ell}\mathcal{U}_{G(\ell)}}_{\gammain(\gauge)}\underbrace{\prod_{\xarg}\fid(\xarg)\ket{\Omega_v}}_{D}.
  \label{eq:def_covariance_matrices}
\end{align}
For further details about the formulation of Gaussian operators in terms of covariance matrices, see Appendix~\ref{sec:app_gaussian}.
The covariance matrices or parts of them allow the efficient calculation of the Monte Carlo transition probability [cmp. Eq.~\eqref{eq:norm_sq}].

\subsection{Construction with multiple layers}
Although the ansatz wave function with a single layer, i.e. two variational parameters, captures the high coupling regime very well, the low coupling regime is challenging for a single layer (cmp. Fig.~\ref{fig:convergence}).
Upon increasing the number of layers, the agreement between exact diagonalization data and the variational PEPS approach improves dramatically.
In order to increase the number of variational parameters, we add more virtual fermions to the construction.
Each layer carries an independent set of parameters, i.e. the matrix $T$ in the fiducial operator $\fid$ is different for each layer, while the states are coupled to the same gauge field.
This ensures that all states fulfill the Gauss law.
The virtual fermions of different layers on the links do not interact.
The complexity of the computation scales linearly in the number of layers because the state can be contracted as independent layers of equally sized PEPSs.
Further details about the contraction and the changes to the calculation of observables are explained in Appendix~\ref{sec:app_layers}.

\section{Computational Evaluation\label{sec:comp_evaluation}}
The ansatz defined above characterizes a family of states that depends on two parameters.
In order to find the ground state of the Hamiltonian~\eqref{eq:hamilton_ZN} for $N=3$, we have to adapt the parameters such that the energy is minimized.
By computing expectation values of observables and derivatives with respect to the parameters via sampling, we circumvent the unfavorable scaling of PEPS contractions.
The variational Monte Carlo technique works in a two step procedure: first, the energy and the gradients are sampled for a given set of parameters $\alpha$.
In the second step, the parameters are changed $\alpha\to\alpha'$ according to the gradients and a minimization algorithm.

\subsection{Calculation of expectation values}
The Hamiltonian~\eqref{eq:hamilton_ZN} consists of two terms, the electric energy and the magnetic energy.
Due to translational invariance of the states and the Hamiltonian, it is sufficient to calculate the energy of a single plaquette and a single link,
\begin{align}
  \expval{H}=&n_{\text{links}}\left(2-\expval{P_\ell+P_\ell\dgr}\right)+\nonumber\\
  &+n_{\text{plaq}}\left(2-\expval{Q_{p_1} Q_{p_2}Q_{p_3}\dgr Q_{p_4}\dgr+\text{H.c.}}\right),
  \label{eq:energy_trans_inv}
\end{align}
where $\nplaq=L^2$, $\nlinks=2\nplaq$ and $L$ is the linear extent of the quadratic lattice (number of vertices).
In the equation above, $\ell$ is a freely chosen link. 
If not stated otherwise, we choose the link at $\xarg=0$ in the horizontal direction.
Calculating the magnetic energy is a special case of the expectation value of a Wilson loop.
We define the Wilson loop operator as 
\begin{align}
  W(R_1,R_2)=\prod_{{\ell}\in C}Q_\ell,
  \label{eq:wilson_loop}
\end{align}
where $C$ is an oriented, rectangular curve of length $R_1$ in the horizontal and $R_2$ in the vertical direction.
\begin{figure}[h]
  \centering
  \includegraphics{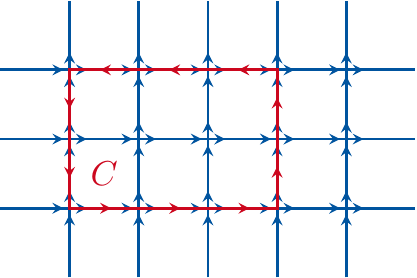}
  \caption{Illustration of a Wilson loop.
    The operator $Q_\ell$ is chosen as is if the red path follows the direction of the blue arrows and daggered if it traverses the blue arrows in the opposite direction.
  }
  \label{fig:wilson_loop}
\end{figure}
The operator $Q_\ell$ is picked as is or daggered according to whether the link is traversed in the direction of the blue arrows (cmp.~Fig.~\ref{fig:wilson_loop}) or against them.
The Wilson loop operator does not only play a role for the calculation of the energy, but can be used as an indicator for confinement in the theory (cmp. Sec.~\ref{sec:results}).
Given the state defined in~\eqref{eq:ansatz_state}, the expectation value of a Wilson loop reads
\begin{align}
  \label{eq:exp_obs_wilson}
  \expval{W(R_1,R_2)}&=\sum_{\mathcal{G}} \mathcal{F}_{W(R_1,R_2)}(\gauge)p(\gauge)\\
  &=\expval{\mathcal{F}_{W(R_1,R_2)}}_{\text{MC}}\nonumber,
\end{align}
where the estimator $\mathcal{F}_{W(R_1,R_2)}=\prod_{\ell\in C}\exp(\pm i\phi(\ell))$ is a complex number and the sampling probability is 
\begin{align}
  p(\gauge)=\frac{\normsq}{\sum_{\gauge'}\abs{\Psi(\gauge')}^2}.
  \label{eq:def_prob}
\end{align}
While the expression $\expval{\cdot}$ is the expectation value of an operator, the expression $\expval{\cdot}_{\text{MC}}$ is a $p(\gauge)$-weighted average over complex numbers.
Since the norm of a state is always real and larger than zero, this formulation of a Monte Carlo procedure cannot suffer from the sign problem.

Using the covariance matrices defined in~\eqref{eq:def_covariance_matrices} in the formulation of Majorana fermions (cmp.~Appendix~\ref{sec:app_gaussian}), we can write the squared norm of the wave function as
\begin{align}
  \normsq=\sqrt{\det(\frac{1-\gammain(\gauge) D}{2})}.
  \label{eq:norm_sq}
\end{align}
It serves as the transition probability between different configuration states of the gauge field.

In our Monte Carlo scheme, we use the Metropolis algorithm~\cite{metropolis_equation_1953} with Eq.~\eqref{eq:def_prob} as a transition probability.
In each step, one gauge field is randomly selected and updated according to the transition probability.
The gauge field is initialized with state $\ket{0}$ everywhere and warmed up without measurements for a fixed number of iterations.
After the warm-up phase, each iteration includes a measurement of the observables.

The electric energy is not diagonal in the gauge field basis.
Instead of evaluating the full electric energy, we focus on the expectation value $\expval{P_{\ell}}$.
$P_{\ell}$ acts as a lowering operator on the gauge field states.
Thus, we have to evaluate an expression that has a modified gauge field on one link.
We can transfer that modification to the covariance matrices by evaluating the integrals in Grassmann variables directly.
The estimator for $\expval{P_\ell}$ in a $\Z{3}$ gauge theory is
\begin{align}
  \mathcal{F}_\text{el}(G)=\frac{1}{4}\frac{\Pf\left(\gammainmod-D\inv\right)}{\sqrt{\det\left(D\inv-\gammain\right)}},
  \label{eq:estimator_el_energy_expression}
\end{align}
where $\gammainmod$ is a modified version of $\gammain$ that differs from the original one on link $\ell$.
Details about the calculation are provided in Appendix~\ref{sec:app_electric_energy}.

\subsection{Evaluation of gradients}
The evaluation of gradients with respect to the parameters in $T$ enables the efficient minimization of observables.
Instead of directly tracking the derivative of the parameters through the state construction, we derive the matrix equations obtained for the covariance matrices with respect to the variational parameters.
The covariance matrix of the fiducial state $D$ does not change during the Monte Carlo computation and is the only one that contains variational parameters $\alpha\in\{y,z\}$.
Thus, we can calculate the gradient for an arbitrary observable $O$ whose estimator $\mathcal{F}_O(D)$ may depend on the covariance matrix $D$ of the fiducial operator explicitly:
\begin{align}
  &\pdv{\alpha}\expval{O}=\pdv{\alpha}\expval{\mathcal{F}_O(D)}_{\text{MC}}\nonumber\\
  &=\expval{\pdv{\alpha}\mathcal{F}_{O}(D)}_{\text{MC}}+\expval{\mathcal{F}_O(D)\frac{\pdv{\alpha}\normsq}{\normsq}}_{\text{MC}}\nonumber\\
  &\phantom{=}-\expval{\mathcal{F}_O(D)}_{\text{MC}}\expval{\frac{\pdv{\alpha}\normsq}{\normsq}}_{\text{MC}}.
  \label{eq:deriv_observable}
\end{align}
Since we are interested in finding the best ground state approximation with our ansatz, we calculate the gradients of the energy.
They consist of two parts, the gradient of the magnetic and the gradient of the electric energy.
In the case of the magnetic energy, the first term on the right-hand side of~\eqref{eq:deriv_observable} vanishes since the gauge field has no explicit dependence on the parameters.
It remains to calculate the expression $\pdv{\alpha}\normsq$ since we know the form of $\normsq$ from the evaluation of the transition probability~\eqref{eq:def_prob} already.
Using Jacobi's formula
\begin{equation}
    \dv{\alpha} \det A(\alpha)=\Tr\left( \Adj(A(\alpha)) \dv{A(\alpha)}{\alpha}\right),
\end{equation}
 we obtain
\begin{align}
  \pdv{\alpha}\normsq&=\pdv{\alpha}\sqrt{\det(\frac{1-\gammain(\gauge) D}{2})}\nonumber \\
  &=-\frac{1}{2^{N+1}} \sqrt{\det(1-\gammain(\gauge) D)}\nonumber\\
  &\phantom{=}\times\Tr(\gammain(\gauge)\pdv{D}{\alpha} (1-\gammain(\gauge) D)^{-1}).
  \label{eq:deriv_norm}
\end{align}
Combining~\eqref{eq:norm_sq} and~\eqref{eq:deriv_norm}, we find 
\begin{align}
  \frac{\pdv{\alpha}\normsq}{\normsq}&=\frac{\pdv{\alpha}\normsq}{\sqrt{\det\left(\frac{1-\gammain(\gauge) D}{2}\right)}}\nonumber\\
  &=-\frac{1}{2}\Tr\left(\gammain(\gauge)\pdv{D}{\alpha}\left(1 - \gammain(\gauge) D\right)^{-1}\right),
\end{align}
where $\pdv{D}{\alpha}$ is the explicit derivative of the covariance matrix of the virtual modes with respect to parameter $\alpha$.
This expression can be derived analytically.

In contrast to the magnetic energy, the electric energy depends explicitly on the parameters of the ansatz.
Thus, the first term on the right-hand side of~\eqref{eq:deriv_observable} does not vanish.
The explicit form of the gradient is stated in Appendix~\ref{sec:app_electric_energy}.

\subsection{Variational minimization}

For small systems ($L=2$), we can substitute the Monte Carlo step with an exact contraction (EC) of the PEPS.
Each possible gauge field configuration on the lattice is sampled and the individual contributions of the different states are summed up.
In the case of exact calculations of the gradients and observables, we used the Broyden-Fletcher-Goldfarb-Shanno (BFGS) algorithm~\cite{press_numerical_2007} to adapt the parameters of the state.
If the gradients and the observables are calculated with Monte Carlo sampling, the inherent error of the estimates makes the use of a line search based algorithm like BFGS difficult.
The fluctuations of the estimate lead to inconsistencies during the line-search which cause the termination of the algorithm.
Thus, we decided to work with a simple gradient descent algorithm if the expectation values are estimated with Monte Carlo.
After estimating the energy and the gradients, we adapt the set of parameters in the opposite direction of the gradient,
\begin{align}
  \alpha'=\alpha-\xi(i)\pdv{\expval{H}}{\alpha}
  \label{eq:grad_descent}
\end{align}
where $\xi(i)$ is the weight for the gradient in dependence of the step.
We used $\xi(i)=0.01\cdot 0.99^i$ in our simulation.
The choice of parameters and the schedule of $\xi(i)$ may be further optimized.

\section{\label{sec:results}Results}
\begin{figure}
  \centering
  \includegraphics[width=\columnwidth]{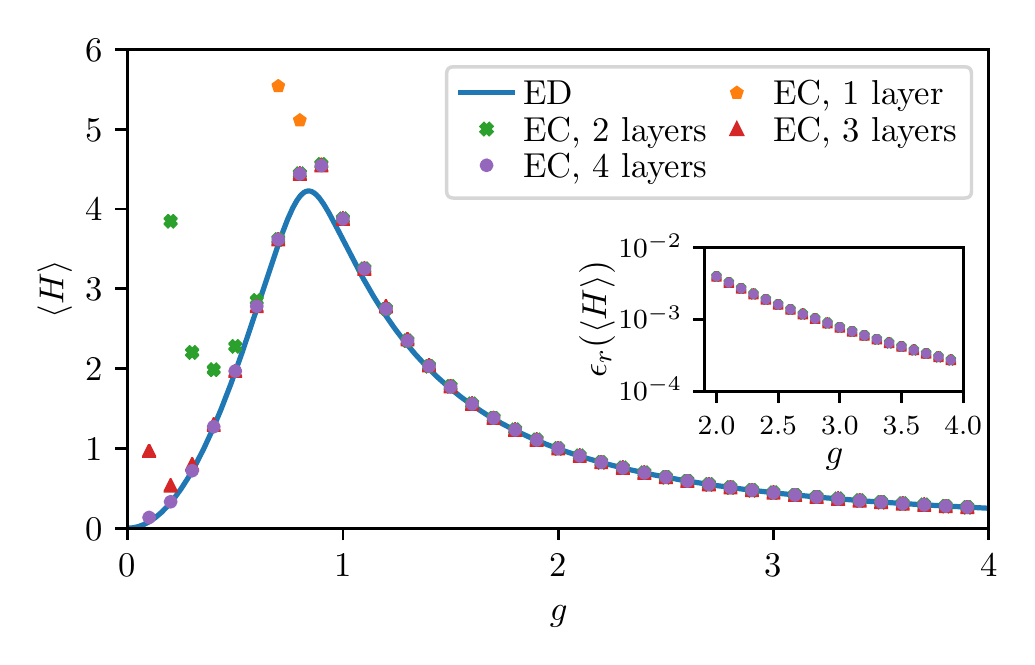}
  \caption{Convergence of the energy for a $L=2$ system. The solid blue line is the exact diagonalization (ED) result.
  The colored dots are exact contractions (ECs) of the ansatz state with varying number of layers of virtual fermions on the links.
  The inset displays the relative error $\epsilon_r$ of the energy with respect to the exact diagonalization results at high coupling.}
  \label{fig:convergence}
\end{figure}
Applying the ansatz developed in Ref.~\cite{zohar_combining_2018} to a physical Hamiltonian, we want to ensure that we are able to capture relevant physics despite the small number of parameters of the states.
In particular, we want to demonstrate that a higher number of layers leads to an improved expressibility.

As a first step, we compare to a small system with $L=2$, i.e. four plaquettes, which can be solved with exact diagonalization (cmp. Fig.~\ref{fig:convergence}).
Due to the small lattice size, we can contract the GGPEPS exactly and do not have to use Monte Carlo.
The figure and the inset show good agreement for states at high couplings where the electric energy is the dominant contribution in the Hamiltonian~\eqref{eq:hamilton_ZN}.
The ground state of the electric Hamiltonian is the state with no electric excitations, i.e. the electric field is zero on all links.
We expect to approximate it well because it is the state that we obtain if the operator $\fid$ is equal to the identity.
This happens if both parameters $y=z=0$: $T(y=0,z=0)=\mathds{1}$.
We observed that the values of $y$ and $z$ approach zero as the coupling increases.

While the high coupling regime matches well to the exact values, the low coupling regime, which is dominated by the magnetic energy, is more challenging.
States with few layers show a divergent behavior at low couplings.
The quadratic divergence is caused by a lack of expressibility of states with few layers:
The parameters approach a constant for low coupling and the $1/g^2$ term in the Hamiltonian leads to the divergence.
An increase in the number of layers helps to systematically improve the states while only linearly affecting the run-time.

The error around the transition $g\approx 1$ does not decrease when additional layers are used.
We attribute this behavior to the specific ansatz that we are using.
We do not expect a Gaussian PEPS based ansatz to hold at criticality.

\begin{figure}
    \centering
    \includegraphics[width=\columnwidth]{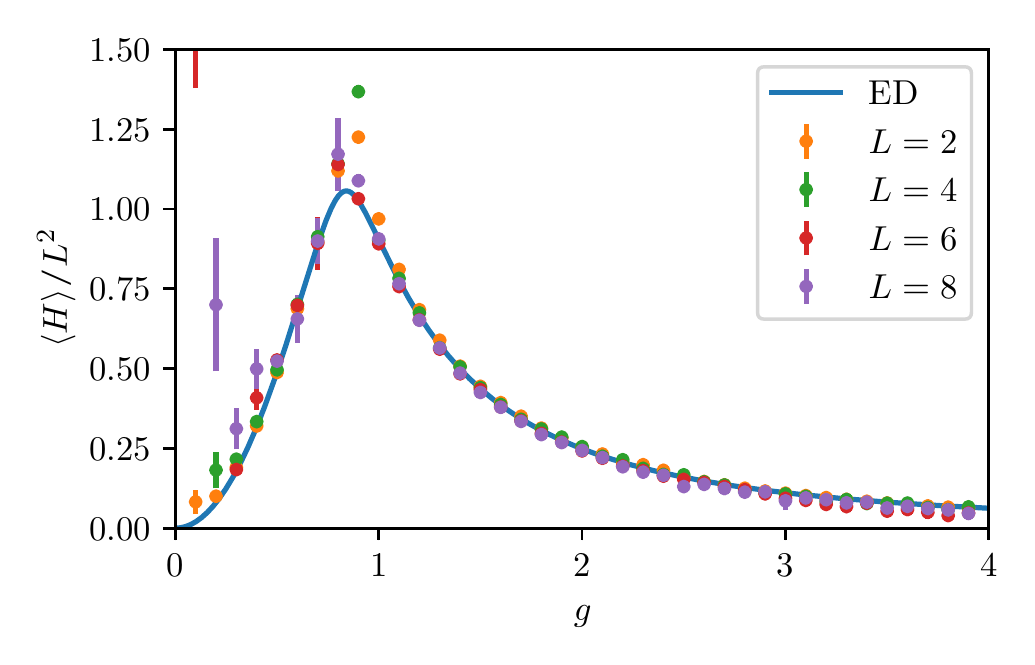}
    \caption{Finite size effects for different system sizes.
    The blue line is the exact data for an $L=2$ system. 
    All data points are computed with VMC for different system sizes using with three layers in the construction of the state.
    }
    \label{fig:finite_size}
\end{figure}
Figure~\ref{fig:finite_size} shows the energy density of the system for different lattice sizes for three layers of the parameters.
Due to the larger system sizes, we cannot contract the GGPEPS exactly.
The Monte Carlo estimation uses $10^4$ steps for the warm-up phase that is performed without measurement and $10^5$ steps for the sampling.
Since the Monte Carlo has to be performed for each variational minimization step, the number of Monte Carlo steps with measurements is kept rather small.
Especially the calculation of the electric energy, which features a Pfaffian, is expensive.

The estimates agree very well with the ED data for an $L=2$ system over a large range of the coupling.
The deviations at the phase transition due to the ansatz as described above.
The deviation at very low coupling for large system sizes originates from the fact that the minimization becomes increasingly costly.
Especially the calculation of the Pfaffian in the electric energy is computationally expensive.
While all determinants that appear in the calculation of norms can be calculated by updating previous results if the gauge field is changed, the Pfaffian has to be recalculated in every step.
The Pfaffian is the single most expensive step in the algorithm.
Since we are plotting the energy density in relation to a $L=2$ system, deviations can be either finite size effects (in which case the MC points would be more correct than ED) or errors due to the Monte Carlo sampling procedure.

Following previous works, we expect the theory to have two phases~\cite{horn_hamiltonian_1979, kogut_introduction_1979}.
According to Elitzur's theorem~\cite{elitzur_impossibility_1975}, the expectation value of any operator that is not gauge invariant will vanish, and thus a local order parameter is ruled out.
Instead, following Wegner and Wilson~\cite{wegner_duality_1971,wilson_confinement_1974}, we can analyze the correlation in the different phases by studying the Wilson loop. 
The corresponding operator is gauge invariant and shows different scaling in the different phases of $\Z{N}$ theories.
In the low-coupling regime, which is dominated by the magnetic part $H_B$ of the Hamiltonian, the expectation value of the Wilson loop follows a perimeter law which, to lowest order in perturbation theory~\cite{kogut_introduction_1979}, reads
\begin{align}
  \expval{W(R_1,R_2)}\sim\exp(-\kappa_p 2(R_1+R_2)).
  \label{eq:wilson_perimeter}
\end{align}
Here, $\kappa_p$ is a constant and $2(R_1+R_2)$ is the perimeter of the Wilson loop.
The scaling changes in the high coupling regime, where
the electric energy is the dominant contribution to the total energy and the Wilson loop operator scales with the area of the curve.
The area scaling reads to lowest order in perturbation theory~\cite{kogut_introduction_1979},
\begin{align}
  \expval{W(R_1,R_2)}\sim \exp(- \sigma R_1R_2),
  \label{eq:wilson_area}
\end{align}
where $\sigma$ is the string tension.
Since the potential of static charges, i.e. charges that are not dynamically coupled to the gauge fields in the Hamiltonian, increases linearly with the distance in this phase, it costs an infinite amount of energy to separate two static charges.
The two static charges are confined.

\begin{figure}
  \centering
  \includegraphics[width=\columnwidth]{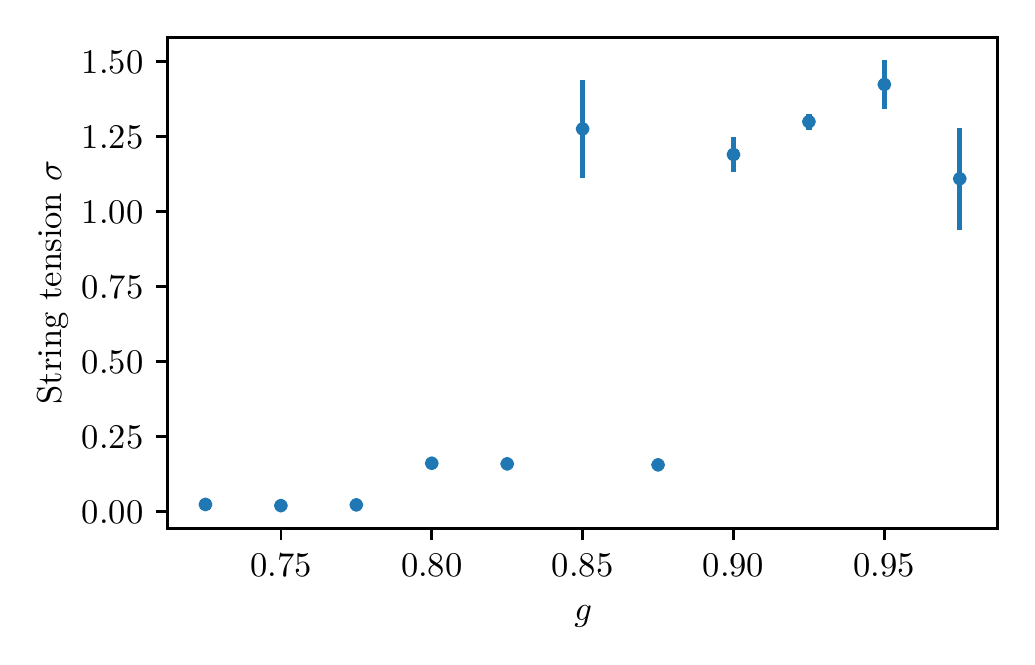}
  \caption{String tension for different value of the coupling.
  The string tension is extracted by fitting the area law expectation to Wilson loops of different size.
  The state is constructed with three layers of virtual fermions.}
  \label{fig:wilson_scaling}
\end{figure}

We can use the states that we obtained using the VMC procedure for an $L=6$ lattice to evaluate the scaling behavior in the different regimes (cmp. Fig.~\ref{fig:wilson_scaling}).
As before, we used three layers in the minimization. 
The Wilson loop expectation values are recomputed for the minimal parameters with $10^4$ warm-up steps and $10^6$ sampling steps.
By fitting~\eqref{eq:wilson_area} to different Wilson loops $W(R_1,R_2)$ of a maximal size of $L/2$ and $|R_1-R_2|<1$, we can obtain the string tension of the states.
The result of the fits for different couplings is shown in Fig.~\ref{fig:wilson_scaling}.
The $\Z{3}$ gauge theory can be mapped to a three state Potts model~\cite{bhanot_phase_1980} and the first order phase transition has been studied with Monte Carlo~\cite{blote_first-order_1979}.
The plot shows that the string tension is almost zero in the low-coupling phase and rises to a finite value in the high-coupling, confining phase.
Around the transition region, the minimization becomes difficult due to the Ansatz we are using.
Thus, results in direct vicinity to the transition region might not be obtained for the ground state and one has to be careful to use them for an interpretation of confining or nonconfining behavior~\cite{polyakov_quark_1977}.
The range of accessible couplings is limited from above since the Wilson loop decays exponentially with size and coupling.
The Monte Carlo procedure cannot reliably resolve the expectation value of the Wilson loop in the high coupling regime.

\section{Conclusion\label{sec:conclusion}}
We show that GGPEPSs are promising ansatz states for $\Z{N}$ lattice gauge theories in two spatial dimensions.
Since the transition probability between two configurations of the gauge field is given by the squared norm of a state, the sign problem is avoided.
The norm as well as the gradients for a given set of parameters can be efficiently computed with the covariance matrix formalism leading to a scalable algorithm.

By contracting small systems exactly we show that the states themselves capture the relevant physics well although they are based only on a small number of parameters.
We demonstrate a systematic improvement of the energy by increasing the number of virtual fermions on the links while impacting the run-time only linearly.

The variational optimization with Monte Carlo is very successful for large couplings, but gets increasingly difficult for smaller couplings and larger lattices.
In this regime, the states have to approximate states dominated by the magnetic interaction in the Hamiltonian.
Since the ansatz is based on the electric vacuum on the links, this regime is challenging.
Additionally, larger lattices lead to higher run-times, especially in the calculation of the Pfaffian in the electric energy.

We expect to be able to improve the results of the Monte Carlo simulation further by changing to a more advanced sampling scheme. 
Currently, the algorithm updates only one spin at a time, which leads to a smaller relative change if the system size increases.
The usage of collective cluster updates~\cite{wolff_collective_1989,swendsen_nonuniversal_1987} or hybrid Monte Carlo techniques~\cite{duane_hybrid_1987} may lead to better convergence. 

Additionally, the ansatz introduced in Ref.~\cite{zohar_combining_2018} allows for static charges and dynamic fermions.
The introduction of static charges allows to measure the string tension directly as an observable between two opposite charges and leads to another measure of confinement which is especially beneficial at large couplings.
Simulating dynamic fermions presents the interesting possibility to study the behavior of mesonic strings. 

Finally, the optimization in the weak coupling regime could be improved by starting from a different initial state on the links.
If the state on the links is more suited for the magnetic Hamiltonian, the physics of the magnetic phase might be easier to capture with fewer layers.

\appendix
\section{\label{sec:app_tmat} Derivation of $T$}
The fiducial operator~\eqref{eq:def_fid_op} used in the GGPEPS construction~\eqref{eq:fermionic_state_gauged} determines the symmetries of the state~$\ket{\Psi}$.
We demand rotational invariance by $\pi/2$, translational invariance when shifting by two sites due to the staggering and charge conjugation invariance if we shift by one site.
Since the parametrization was originally developed to accommodate a $U(1)$ gauge theory~\cite{zohar_fermionic_2015}, the formulation obeys, additionally, a global $U(1)$ symmetry.
Here, we state only the result
\newcommand{\sqrtz}{z/\sqrt{2}}
\begin{align}
  T=\mqty(0&&y&&\sqrtz&&\sqrtz\\-y&&0&&-\sqrtz&&\sqrtz\\-\sqrtz&&\sqrtz&&0&&y\\-\sqrtz&&-\sqrtz&&-y&&0),
  \label{eq:def_t_matrix}
\end{align}
with $y,z\in\mathds{C}$.
$y$ and $z$ are the only two independent parameters that remain. 
The matrix is given in the mode order $\{l,r,u,d\}$.
The rows correspond to the modes $\{l_+,r_-,u_-,d_+\}$, and the columns to $\{l_-,r_+,u_+,d_-\}$.
In this work, we restrict ourselves to $y,z\in\mathds{R}$.

\section{\label{sec:app_layers} Formalism with multiple layers}
We achieve a higher expressibility of the ansatz states by increasing the number of virtual fermions on the links.
Different layers of virtual fermions do not interact with each other and have independent sets of parameters $y^{(i)}$ and $z^{(i)}$, where $i$ is the index of the layer.
They can be seen as different PEPSs coupled to the same gauge field.
Thus, the norm of the state $\ket{\Psi}$ is the product of the norms of its layers $\ket{\Psi_i}$:
\begin{align}
    \braket{\Psi}=\prod_i \braket{\Psi_i},
\end{align}
where $i$ is the index of the layer and runs from 1 to the number of layers.
This construction leads to a linear scaling with the bond dimension.
The matrix size of the covariance matrices stays unchanged because we do not add the parameters to the $T$ matrix.
Instead, we consider multiple covariance matrices generated by different matrices $T_i$.
Thus, we have to perform parts of the calculation multiple times with varying covariance matrices of the same size.

Since we layer only the virtual fermions, the computation of diagonal observables in the gauge field does not change.
Observables like the electric energy, however, need more consideration.
Due to the product structure of the ansatz state, we can write the estimator of the electric energy as a product $\mathcal{F}_{\text{el}}=\prod_i \mathcal{F}_{\text{el}}^{(i)}$, where $i$ is again the index of the layer.
Each $\mathcal{F}_{\text{el}}^{(i)}$ involves only the covariance matrices of layer $i$ and can be calculated with Eq.~\eqref{eq:app_num_electric_energy}.

Finally, the gradients for the squared norm and the explicit derivative of the electric energy have to be adapted.
The derivative of the squared norm enters the equations only as a fraction of the squared norm [cmp.~Eq.~\eqref{eq:deriv_observable}], we only have to adapt the expression
\begin{align}
  &\frac{\pdv{\alpha_i}\prod_j \braket{\Psi_j(\gauge)}}{\prod_j\braket{\Psi_j(\gauge)}}\nonumber\\
  &=\frac{\sum_i\prod_{i\neq j}\braket{\Psi_j(\gauge)}\pdv{\alpha_i}\braket{\Psi_i(\gauge)}}{\prod_{j}\braket{\Psi_j(\gauge)}}\nonumber\\
  &=\frac{\pdv{\alpha_i}\braket{\Psi_i(\gauge)}}{\braket{\Psi_i(\gauge)}}.
\end{align}
Here, we move the derivative with respect to parameter $\alpha_i\in\{y,z\}$ of layer $i$ to the respective layer $i$ since all other parameters are independent of $\alpha_i$.

The gradient of the electric energy is adapted in a similar fashion because the derivative acts only on one of the layers.

\section{\label{sec:app_gaussian}Gaussian formalism}
Given a Dirac mode $c$, we can construct the corresponding Majorana operators $\gamma^{(1)}$ and $\gamma^{(2)}$ as
\begin{align}
  \gamma^{(1)}&=c+c^{\dagger}\nonumber\\
  \gamma^{(2)}&=i(c-c^{\dagger}).
  \label{eq:def_majorana}
\end{align}
The Majorana modes obey the anticommutation relation $\anticommutator{\gamma_a}{\gamma_b}=2\delta_{a,b}$.
The construction~\eqref{eq:fermionic_state_gauged} uses only Gaussian operators, thus, we can formulate it in terms of covariance matrices.
We define the covariance matrix of a Gaussian state $\ket{\Phi}$ in terms of Majorana modes as
\begin{align}
    \Gamma_{a,b}=\frac{i}{2}\expval{\comm{\gamma_a}{\gamma_b}}=\frac{i}{2}\frac{\bra{\Phi}\comm{\gamma_a}{\gamma_b}\ket{\Phi}}{\braket{\Phi}}.
\end{align}

The construction of the Gaussian state is divided into two covariance matrices.
We separate the covariance matrix of the fiducial operators $D$ from the covariance matrix of the gauged projectors $\gammain(\gauge)$.
This allows us to calculate the squared norm of the state with Eq.~\eqref{eq:identities_bravyi}.
During one Monte Carlo run, $D$ stays constant and can be calculated during the initialization.
Changing the gauge field value on a link only alters $\gammain(\gauge)$.
We refer to Ref.~\cite{zohar_combining_2018} for more details on the Gaussian mapping.

In order to calculate the squared norm of the wave function, we use the following identities~\cite{bravyi_lagrangian_2005}:
\begin{align}
  &\int D\theta \exp(\frac{i}{2}\theta^{T}M\theta)=i^{n}\Pf \left(M\right)\nonumber\\
  &\begin{alignedat}{1}
    \int D\theta \exp(\eta^{T}\theta+\frac{i}{2}\theta^{T}M\theta)=i^{n}&\Pf\left(M\right)\times\nonumber\\
    &\times\exp(-\frac{i}{2}\eta^{T}M^{-1}\eta)\nonumber
  \end{alignedat}\nonumber\\
  &\Tr(XY)=(-2)^{n}\int D\theta D\mu e^{\theta^{T}\mu}[X]_{G,\theta}[Y]_{G,\mu}, 
  \label{eq:identities_bravyi}
\end{align}
where M is a complex antisymmetric $2n\times 2n$ matrix and $[X]_{G,\theta}$ is the Grassmann representation of the operator $X$ in terms of Grassmann variables $\theta$.
Equation~\eqref{eq:norm_sq} follows directly from~\eqref{eq:identities_bravyi}.

\section{\label{sec:app_electric_energy}Calculation of the electric energy and its gradient for \Z{N}}
\subsection{Calculation of the expectation value of the electric energy}
Since the electric energy is not diagonal in group element basis, we cannot use the equivalent of~\eqref{eq:exp_obs_wilson} directly.
Due to the translational invariance of the states and the Hamiltonian, it is sufficient to calculate the expectation value of the electric energy over one link $\ell$.
The notation for $\Psi(\gauge)$ introduced in~\eqref{eq:fermionic_state_gauged} is changed to distinguish between the group element $q$ on link $\ell$ and all other group elements $G$ to $\Psi(q,G)$.
In the following, we focus on the calculation of the expectation value $\expval{P_\ell}$; the extension to $\expval{P_\ell+P_\ell\dgr}$ which appears in the Hamiltonian\eqref{eq:hamilton_ZN} follows directly.
Since we are only considering a single, fixed link for the rest of the calculation, we drop the index $\ell$:
\begin{align}
  \expval{P}\nonumber &=\frac{\bra{\Psi}P\ket{\Psi}}{\braket{\Psi}}\nonumber\\
  &=\sum_{q,q',G}\bra{q'}P\ket{q}\frac{\Psi^*(G,q')\Psi(G,q)}{\normsq}p(G,q)\nonumber\\
  &=\sum_{q,G}\frac{\Psi^*(G,q-1)\Psi(G,q)}{\normsq}p(G,q)\nonumber\\
  &=\sum_{q,G}\mathcal{F}_\text{el}(G,q)p(G,q),
  \label{eq:app_el_en_mc_estimator}
\end{align}
where $\mathcal{F}_\text{el}(G)$ is the Monte Carlo estimator of the electric energy.
From the second line to the third line we use that $P$ acts as a lowering operator on the gauge field states.
The remaining expression is the product of two wave functions that differ in terms of the gauge field on one link. Using the explicit formulation of the state, we obtain [product symbols as in~\eqref{eq:fermionic_state_gauged}]
\begin{align}
&\Psi^*(G,q')\Psi(G,q)\nonumber\\
&=\bra{\Omega_v}\fid\dgr\mathcal{U}_{(q',G)}\dgr\omega\mathcal{U}_{(q,G)}\fid\ket{\Omega_v}\nonumber\\
&=\bra{\Omega_v}\fid\dgr\mathcal{U}_{(q,G)}\dgr\mathcal{U}_{(\tilde{q})}\omega\mathcal{U}_{(q,G)}\fid\ket{\Omega_v}.
\end{align}
Thus, we calculate the expectation value of the new operator $\mathcal{U}_{(\tilde{q})}\omega$ with the density matrix resulting from the original wave function $\Psi(\mathcal{G})$.
Since we gauge only the right and upper modes, we can focus on the gauging transformation $\mathcal{U}_{(q')}=\exp(i\Phi r_{+}\dgr r_{+})=\exp(i\Phi \rd r)$ with $\Phi=\pm\delta$.
Without loss of generality, we choose a right mode for the computation.
We consider only positive modes $r_+$ for simplicity.
The negative modes $r_{-}$ are gauged with the same expression where $\Phi$ is substituted by $-\Phi$.
For increased readability, we will skip the plus and minus signs of the modes in the following calculation:
\begin{align*}
  \mathcal{U}_{(\tilde{q})}\omega&= e^{i\Phi\rd r}(1+\ld l) r\rd l\ld(1+lr)\\
  &=rl+r\rd l\ld+e^{i\Phi}\ld l\rd r+e^{i\Phi}\ld\rd
\end{align*}
We use the Majorana modes~\eqref{eq:def_majorana} to rewrite $\mathcal{U}_{(\tilde{q})}\omega$ with $p=1+e^{i\Phi}$ and $m=e^{i\Phi}-1$:
\begin{align}
  \mathcal{U}_{(\tilde{q})}\omega=&\frac{1}{4}p\left[ 1-\frac{m}{p}r_{1}l_{1}-i r_{1}l_{1} + \frac{m}{p}r_{2}l_{2}- i r_{2}l_{1}\right.\nonumber\\
                                  &\phantom{\frac{1}{4}p[}\left.+ i \frac{m}{p}r_{1}r_{2}+ i\frac{m}{p}l_{1}l_{2} \right]\nonumber\\
                                  &+\frac{1}{4}p\left[ -r_{1}r_{2}l_{1}l_{2} \right].
  \label{eq:app_el_en_majorana}
\end{align}
Following \cite{bravyi_lagrangian_2005}, we replace the Majorana operators with Grassmann variables, to calculate the overlap:
\begin{align}
  [\mathcal{U}_{(\tilde{q})}\omega]_G&=\left(-\frac{m}{p}\theta_{r_{1}}\theta_{l_{1}}\right)\left(\frac{m}{p}\theta_{r_{2}}\theta_{l_{2}}\right)\nonumber\\
  &+\left(-i\theta_{r_{1}}\theta_{l_{2}}\right)\left(-i\theta_{r_{2}}\theta_{l_{1}}\right)\nonumber\\
  &+\left(i\frac{m}{p}\right)^{2}\theta_{r_{1}}\theta_{r_{2}}\theta_{l_{1}}\theta_{l_{2}}.
  \label{eq:app_el_en_grassmann}
\end{align}
Finally, we can formulate~\eqref{eq:app_el_en_grassmann} as a matrix for the full operator $U_{(\tilde{q})}\omega$:
\begin{align}
  &U_{(\tilde{q})}\omega=\frac{1}{4}(1+e^{i\Phi})\nonumber\\
  &\times\exp\left( \frac{i}{2} (\theta_{r_{1}}\,\theta_{r_{2}}\,\theta_{l_{1}}\,\theta_{l_{2}}) \underbrace{\mqty(0&&it&&-t&&-1\\-it&&0&&-1&&t\\t&&1&&0&&it\\1&&-t&&-it&&0)}_{M(\Phi)}\mqty(\theta_{r_{1}}\\\theta_{r_{2}}\\\theta_{l_{1}}\\\theta_{l_{2}})\right),
  \label{eq:electric_field_op_matrix}
\end{align}
where $t=\tan(\frac{\Phi}{2})$.
The covariance matrix $M(\Phi)$ in~\eqref{eq:electric_field_op_matrix} of the $r$ and $l$ modes replaces a part of the original covariance matrix \gammain that belongs to the link that $\mathcal{U}_{(\tilde{q})}$ acts on.
Since one link consists of positive and negative modes, we will have to substitute the single link with the direct sum $M(\Phi)\oplus M(-\Phi)$.

Due to the modification of the original covariance matrix for the projectors, we have to adapt the calculation for the overlap of two wave functions.
While the identities~\eqref{eq:identities_bravyi} still hold, formula~\eqref{eq:norm_sq} cannot be used.
Instead we calculate the overlap using
\begin{align}
  \Tr(XY)=2^{-n}\Pf\left(\Gamma_X\right)\Pf\left(\Gamma_Y-\Gamma_X^{-1}\right)
  \label{eq:overlap_general}
\end{align}
which follows from~\eqref{eq:identities_bravyi}.
Here, $X$ and $Y$ are operators and $\Gamma_X$ and $\Gamma_Y$ are the covariance matrices of $X$ and $Y$ in terms of Grassmann variables.
If the operators are Gaussian, these representations coincide with the covariance matrices in terms of Majorana fermions.

The Grassmann representation of the involved operators is
\begin{align}
  &[\rho]_{G,\mu}=\frac{1}{2^{n}}\exp(\frac{i}{2}\mu^TD\mu)\\
  &[\mathcal{U}_q\dgr\omega]_{G,\theta}\nonumber\\
  &= \frac{1}{2}(1+\cos(\Phi))\frac{1}{2^{n}}\exp\left(\frac{i}{2}\theta^T\left(\bigoplus_{l}^{\nlinks-2\,\text{copies}}\gammain(\ell)\right)\theta\right)\nonumber\\
  &\times\exp(\frac{i}{2}\theta^TM(\Phi)\theta)\exp(\frac{i}{2}\theta^TM(-\Phi)\theta) \label{eq:app_el_en_def_gammamod}.
\end{align}
Here, $\gammain(\ell)$ is the covariance matrix of link $\ell$.
Thus, we have to use an adapted prefactor for~\eqref{eq:overlap_general}:
\begin{align*}
  \Tr(\mathcal{U}_q\dgr\omega\rho)=\frac{1}{2}(1+\cos(\Phi)) 2^{-n}\Pf\left(D\right)\Pf\left(\widetilde{\Gamma}_\text{in}-D^{-1}\right),
\end{align*}
where $\widetilde{\Gamma}_\text{in}$ is the modified covariance matrix of the links as defined in~\eqref{eq:app_el_en_def_gammamod}.
In the case of a \Z{3} gauge, we know that $\cos(\Phi)=-\frac{1}{2}$ and obtain
\begin{align}
  \Tr(\mathcal{U}_g\dgr\omega\rho)=\frac{1}{4} 2^{-n}\Pf\left(D\right)\Pf\left(\widetilde{\Gamma}_\text{in}-D^{-1}\right).
  \label{eq:app_num_electric_energy}
\end{align}
This expression can be further simplified since the Monte Carlo estimator~\eqref{eq:app_el_en_mc_estimator} divides by the square of the norm and we obtain
\begin{align}
  \mathcal{F}_\text{el}(G)=\frac{1}{4}\frac{\Pf\left(\gammainmod-D\inv\right)}{\sqrt{\det\left(D\inv-\gammain\right)}}. \label{eq:app_estimator_el_energy_expression}
\end{align}
This is the expression stated in the main text as Eq.~\eqref{eq:estimator_el_energy_expression}.
In the case of a pure gauge theory,~\eqref{eq:app_estimator_el_energy_expression} can be further simplified with $D\inv=-D$.

\subsection{Calculation of the gradient of the electric energy}
In contrast to the calculation of the gradient of the Wilson loop, we cannot neglect the first term in~\eqref{eq:deriv_observable}.
The estimator of the electric energy depends explicitly on the parameters of the ansatz.
Thus, we have to build the derivative of $\mathcal{F}_{\text{el}}$~\eqref{eq:app_estimator_el_energy_expression}, the estimator of the electric energy, with respect to the parameters $\alpha\in\{y,z\}$.
\newcommand{\sqrtdet}{\sqrt{\det(\mathds{1}-\gammain D)}}
\newcommand{\derivD}{\pdv{D}{\alpha}}
\begin{widetext}
  \begin{align}
    \pdv{\alpha}\mathcal{F}_\text{el}(\gauge,D)
    =\frac{1}{2}\mathcal{F}_\text{el}(\gauge,D) \left[ \Tr(D\inv \derivD)+\Tr(\left(\gammainmod-D\inv\right)\inv D\inv\derivD D\inv)+\Tr(\gammain\derivD D\inv\left(D\inv-\gammain\right)\inv)  \right].
    \label{eq:app_deriv_el_energy}
  \end{align}
\end{widetext}
As above, the expression for $\pdv{D}{\alpha}$ is an analytical expression.
Since $D$ is a covariance matrix of Majorana fermions in a pure gauge theory, $D\inv=D\dgr=-D$ holds.
Thus, the first trace of~\eqref{eq:app_deriv_el_energy} is zero.

\begin{acknowledgments}
Patrick Emonts thanks Julian Bender, Jeanne Colbois, Daniel Robaina, Stefan Wessel and Thorsten B. Wahl for fruitful discussions.
This work was partially funded by the Deutsche Forschungsgemeinschaft (DFG, German Research Foundation) under Germany's Excellence Strategy -- EXC-2111 -- 390814868.
Patrick Emonts acknowledges support from the International Max-Planck Research School for Quantum Science and Technology (IMPRS-QST) as well as support by the EU-QUANTERA project QTFLAG (BMBF Grant No. 13N14780).
P.E. thanks the Hebrew University of Jerusalem for the hospitality during his stay at the Racah Institute of Physics.
\end{acknowledgments}

\bibliography{references.bib}

\end{document}